\documentclass[%
 reprint,
 amsmath,amssymb,
 aps,
]{revtex4-1}

\usepackage{graphicx}
\usepackage{dcolumn}
\usepackage{bm}

\begin{document}


\title{Engineering the ligand states by surface functionalization: A new way to enhance the ferromagnetism of CrI$_3$}

\author{Hongxing Li$^\textit{$^{1}$}$}
\author{Zi-Peng Cheng$^\textit{$^{1}$}$}
\author{Bin-Guang He$^\textit{$^{1}$}$}
\author{Wei-Bing Zhang$^\textit{$^{1}$}$}%
 \email{zhangwb@csust.edu.cn}
\affiliation{%
 $^\textit{$^{1}$}$Hunan Provincial Key Laboratory of Flexible Electronic Materials Genome Engineering, School of Physics and Electronic Sciences, Changsha University of Science and Technology, Changsha 410114, People's Republic of China}%

\date{\today}

\begin{abstract}
The newly discovered 2D magnetic materials provide new opportunities for basic physics and device applications. However, their low Curie temperature (T$_C$) is a common weakness. In this paper, by combining magnetic Hamiltonian, Wannier functions and first-principle calculations, we systematically study the magnetic properties of monolayer CrI$_3$ functionalized by halogen. The magnetic exchange coupling (EX) and magnetic anisotropy (MA) are found to increase significantly by X (X=F, Cl and Br) atom adsorption, and increase along with the coverage of X atom. In the frame work of superexchange theory, the enhanced EX can be ascribed to the reduced energy difference and increased hopping strength between Cr d and I p orbitals, due to the states of I ligand are engineered by X adatom. Besides, the X adatom may provide additional ferromagnetic superexchange channel. Finally, the CrI$_3$ that one side is fully adsorbed by F atoms is found to be a room temperature ferromagnetic semiconductor with T$_C$=650 K. Our results not only give an insightful understanding for the enhancement of ferromagnetism of CrI$_3$ by atom adsorption, but also propose a promising way to improve the ferromagnetism of 2D magnetic materials.
\end{abstract}

\pacs{Valid PACS appear here}
\maketitle


\section{\label{sec:level1}Introduction}
Magnetism in two-dimension (2D) is an interesting topic for several decades, due to the novel physical phenomena and promising applications in low-dimensional devices.\cite{lin2019review,hong2017atomic,li1992dimensional,theorytwo} In 2017, the intrinsic ferromagnetism is discovered for the first time in atomically thin layers of CrI$_3$ and Cr$_2$Ge$_2$Te$_6$,\cite{gong2017discovery,huang2017layer} which provide new platforms for the study of low-dimensional magnetic phenomenon and spintronics device. For example, the photominescence of monolayer CrI$_3$ is spontaneous circularly polarized, in which the helicity is determined by the magnetization direction.\cite{seyler2018ligand} The van der Waals magnetic tunneling junctions that composed by multi-layer CrI$_3$ demonstrate giant tunneling magnetoresistance.\cite{song2018giant,klein2018probing}

Unfortunately, the Curie temperature (T$_C$) is always very low, especially the 2D semiconductor ferromagnet, as a result of its low carrier concentration and few ligands. For instance, T$_C$ of monolayer CrI$_3$ and bilayer Cr$_2$Ge$_2$Te$_6$ are 45 K and 28 K, respectively.\cite{gong2017discovery,huang2017layer} The ultra-low T$_C$ means the sample must be cooled to a very low temperature to maintain the long-ranged magnetic order, which is very expensive and inconvenient. To overcome this obstacle, many strategies, including electrostatic doping and strain, are adopted to improve the T$_C$.\cite{jiang2018controlling,dong2019strain} Especially, Huang \emph{et al}. propose a new general strategy to improve the T$_C$ of CrI$_3$ by isovalent alloying, and they revealed that the enhancement can be ascribed to the reduced virtual exchange.\cite{huang2018toward} Our previous work also demonstrates the ferromagnetism of CrI$_3$ can be enhanced in vdW heterostructure, as a result of interlayer charge transfer.\cite{li2019enhanced}

Since a large part of atoms expose to the surface, the properties of 2D materials are very sensitive to surface functionalization. There are many works about the effects of surface functionalization on the electronic structure and topological properties of 2D materials.\cite{xu2018interfacial,wu2014prediction} Recently, this method has been used to modify the magnetism of 2D magnetic materials. C. Frey\emph{et al}. functionalize MXene surface by different nonmetal elements, finding that the magnetic order and magnetic anisotropy vary with elements. This is because the nonmetal atoms bond with magnetic ions directly, and then the spin-orbit coupling and electron localization of magnetic ions are manipulated.\cite{frey2019surface,frey2018tuning} Upon metal atom adsorption, owing to carriers doping, both magnetic coupling and magnetic anisotropy of CrI$_3$ and Cr$_2$Ge$_2$Te$_6$ can be tuned. However, the carriers doping also transfer the semiconductors to metals, which limit their applications.\cite{guo2018half,kim2019exploitable,song2019tunable}

In the present work, by employing of first-principle calculations, we functionalize monolayer CrI$_3$ by the congeners of I, namely X=F, Cl, Br and I. We find X adatom bonds with I atom of CrI$_3$ strongly. The magnetic exchange coupling (EX) and magnetic anisotropy (MA) of monolayer CrI$_3$ will be enhanced by the adsorption of F, Cl and Br atoms, and increase with the coverage. To understand the underlying mechanism, the magnetic Hamiltonian and Wannier function calculations are carried out. At last, we fully functionalize one side of CrI$_3$ by F atom, finding this system is a room temperature 2D ferromagnetic semiconductor.

\begin{table}
  \caption{The calculated results. \emph{d}$_{I-X}$, the distance between I and adatom, in {\AA}. $\Delta \rho$, the charge transfer from CrI$_3$ to adatom, in e. \emph{d}$_{Cr-I}$, the length of bond Cr-I in which I atom is attached by X adatom, in {\AA}. $\angle_{Cr-I-Cr}$, the bond angle of Cr-I-Cr in which I atom is attached by X atom, in degree. M$_X$ and M$_I$, the magnetic moment of X atom and I atom that bonds with X atom, $\mu_B$. In pristine CrI$_3$, \emph{d}$_{Cr-I}$=2.747 {\AA}, $\angle_{Cr-I-Cr}$=94.77$^{\circ}$, and M$_I$ is -0.081 $\mu_B$.}
  \label{tbl}
  \setlength{\tabcolsep}{3.5mm}{
  \begin{tabular}{cccccc}
    \hline
     \quad & F & Cl & Br & I \\
    \hline
    \emph{d}$_{I-X}$ & 2.20 & 2.76 & 2.93 & 3.07 \\
    $\Delta$$\rho$ & 0.614 & 0.428 & 0.341 & 0.231 \\
    \emph{d}$_{Cr-I}$ & 2.696 & 2.723 & 2.729 & 2.798 \\
    $\angle$$_{Cr-I-Cr}$ & 95.93 & 95.62 & 95.63 & 95.03 \\
    M$_I$ & -0.377 & -0.331 & -0.304 & -0.261 \\
    M$_X$ & -0.290 & -0.374 & -0.405 & -0.396 \\
      \hline
  \end{tabular}}
\end{table}

\section{\label{sec:level1}Calculation method}
All our spin-polarized calculations were performed on Vienna \emph{ab} initio simulation package (VASP) based on density functional theory.\cite{vasp1,vasp2} The ion-electron interactions were described by projector augmented wave (PAW) method.\cite{paw} Perdew-Burke-Ernzerhof (PBE) functional within generalized gradient approximation (GGA) was used to evaluated electron exchange-correlation.\cite{gga} The cutoff energy of plane wave basis was set to 500 eV. The atomic structure was fully relaxed until the residual force on each atom is smaller than 0.01 eV/{\AA} and the convergence criteria of electronic step is 10$^{-5}$ eV. The DFT-D2 method was adopted to account for van der Waals interactions.\cite{d2} A vacuum space of 15 {\AA} was used so that the artificial interactions between the images of slabs can be neglected. Brillouin zone sampling was performed by a 7$\times$7$\times$1 $\Gamma$-center mesh.\cite{kpoint} The Wannier functions calculations were implemented by Wannier90 Code.\cite{wannier90}

\section{\label{sec:level1}Results and disscussions}
\subsection{\label{sec:level2}Single-atom adsorption}

\begin{figure}
\begin{center}
\includegraphics[width=0.5\textwidth]{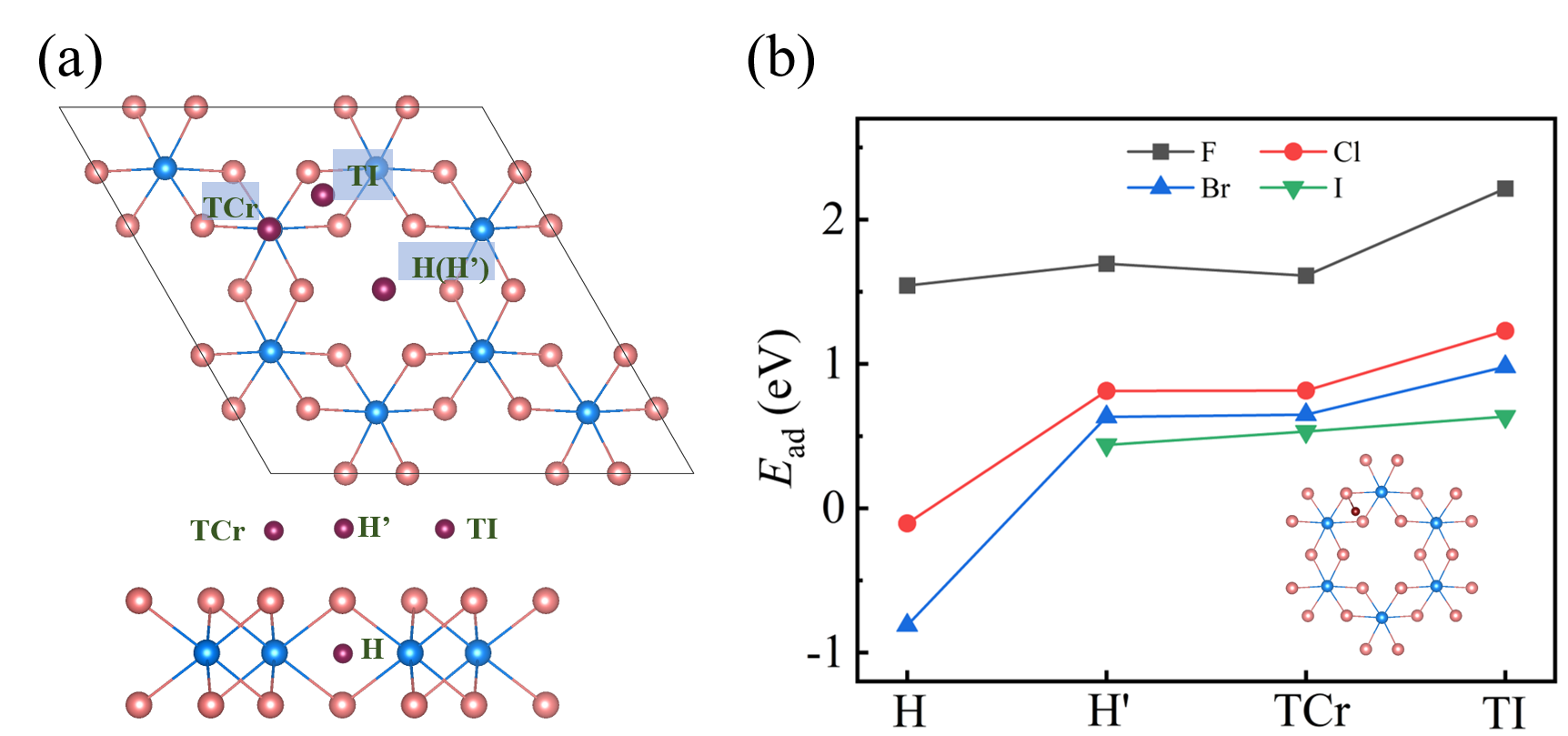}
\caption{(a) The schematic diagram of the adsorption sites on monolayer CrI$_3$, (b) The adsorption energy E$_{ad}$ of X adatom at difference site, the inset is the optimized atomic structure of TI configuration. The pink, blue and purple balls represent the I, Cr and X adatom.}
\label{Fig1}
\end{center}
\end{figure}

The single-atom adsorption is modeled by adsorbing one halogen (X=F, Cl, Br and I) atom on 2$\times$2 supercell of CrI$_3$, and the corresponding coverage is 8.3\% monolayer. The lattice parameter of CrI$_3$ is optimized to be 7.03 {\AA}, so the distance between adjacent adatoms is 14.06 {\AA} and the interaction among them can be neglected. To determine the most stable adsorption configuration, we consider four possible sites on monolayer CrI$_3$, namely the top of Cr atom (TCr), top of I atom (TI), hollow site (H), and top of hollow site (H'), as shown in Fig.~\ref{Fig1}(a). The adsorption energy \emph{E}$_a$ of adatom is calculated by
\begin{equation}
E_{ad}=E_{CrI_3}+E_{X}-E_{tot}
\end{equation}

\noindent Where \emph{E}$_{CrI_3}$, \emph{E}$_X$ and \emph{E}$_{tot}$ are the energies of monolayer CrI$_3$, isolated X atom and the adsorbed system X@CrI$_3$. By this definition, the positive value of \emph{E}$_{ad}$ suggests the adsorption is energetically favorable, while the negative value manifests an energy consumed adsorption. We find that all X atoms have largest positive \emph{E}$_{ad}$ at TI site. However, the largest \emph{E}$_{ad}$ decreases monotonically from 2.216 eV of F to 0.636 eV of I atom, as the reduced electronegative. The results indicate all the X atoms can absorb on CrI$_3$ stably. Bader charge analysis is used to calculate the charge transfer between CrI$_3$ layer and X adatom.\cite{bader} The electron is found to transfer from CrI$_3$ to X atom, which is similar to the situation that X atom adsorbs on MoS$_2$. However, CrI$_3$ transfers more electron to X adatom than MoS$_2$ at the same GGA-PBE level. For example, F adatom gets 0.614 e from CrI$_3$, while it is just 0.572 e from MoS$_2$.\cite{mos2} The large magnitudes of adsorption energy and charge transfer indicate the covalent bonds are formed between I and X atoms.\cite{mos2} Interestingly, the length between Cr atom and the I atom that bonds with X adatom decreases from 2.747 to 2.696 {\AA}, which means the Cr-I bond is strengthened by the adsorption of X atom.

\begin{figure}
\begin{center}
\includegraphics[width=0.43\textwidth]{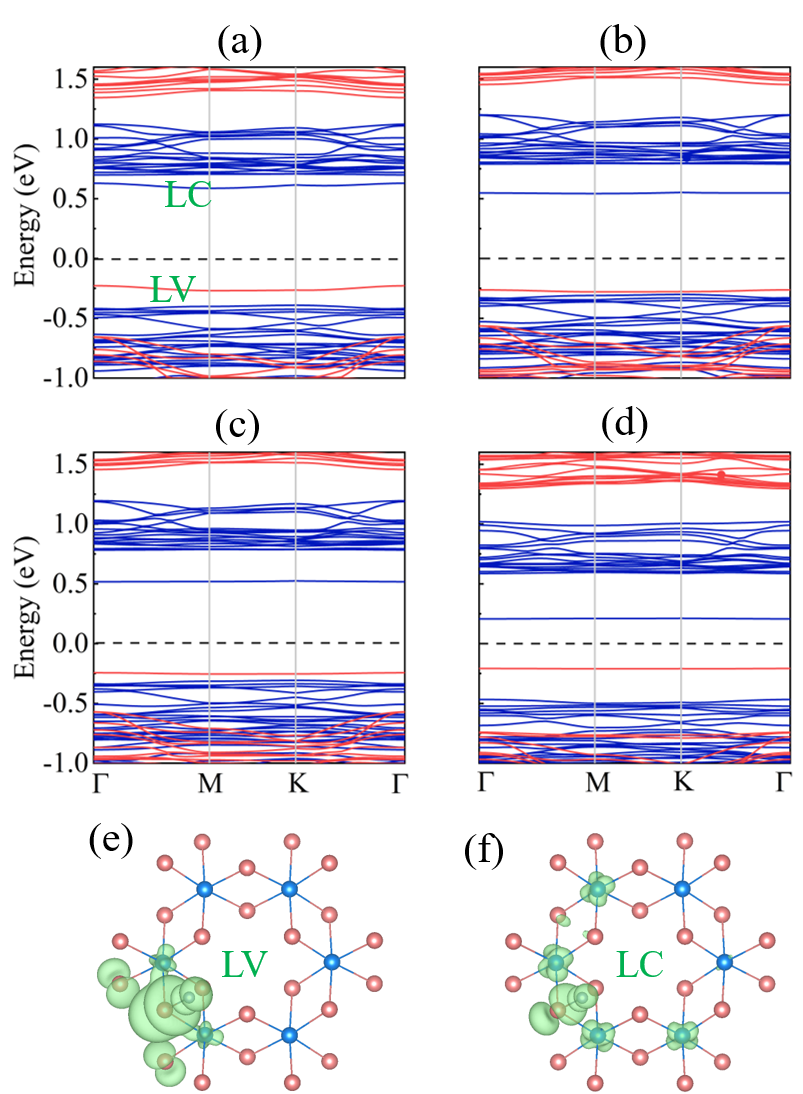}
\caption{(a) (b) (c) and (d), the spin-polarized band structure of  CrI$_3$ adsorbed with F, Cl, Br and I atom, respectively. Blue line is spin-up and red line is spin down. (e) and (f), the partial charge density of band LV and LC as demonstrated in (a). The isosurface is set to 0.001 e/{\AA}$^3$.}
\label{Fig2}
\end{center}
\end{figure}

The spin-polarized band structures are calculated to understand the electronic properties of X@CrI$_3$. As we can see in Fig.~\ref{Fig2}, in the gap region, there are two localized bands. One is close to conduction band (marked as LC), and the other one is near valance band (marked as LV). The distance between LC and LV decreases from F to I. Even so, the systems are still semiconductors. It should be noted that the spin direction of LV is downward, which is opposite to the upward spin of valence band. Fig.~\ref{Fig2}(e) and (f) show the band-decomposed charge density of LV and LC of F@CrI$_3$. We can see that the charge density is distributed around F adatom, Cr and I atoms, indicating the hybridization between the adatom and neighbor atoms. The extension of the isosurface of LV and LC around I atom is larger than F adatom, so the contribution of I atom to the localized band is greater than F adatom. By examining the shape of isosurface, we find LV around Cr atom is t$_{2g}$ feature. Furthermore, two lobes of d orbital connect with the p$_\pi$ orbital of I atom, confirming the bonding state. At the same time, F adatom bonds with I atom by p$_\pi$ hybridization. Different from LV, the LC is mainly contributed by Cr d$_{x^2-y^2}$ orbital. This is accordant with the fact that valance band of monolayer CrI$_3$ is dominate by t$_{2g}$ orbital, while e$_g$ is the primary ingredient of conduction band.\cite{jmcc,lado2017}

\begin{figure}
\begin{center}
\includegraphics[width=0.45\textwidth]{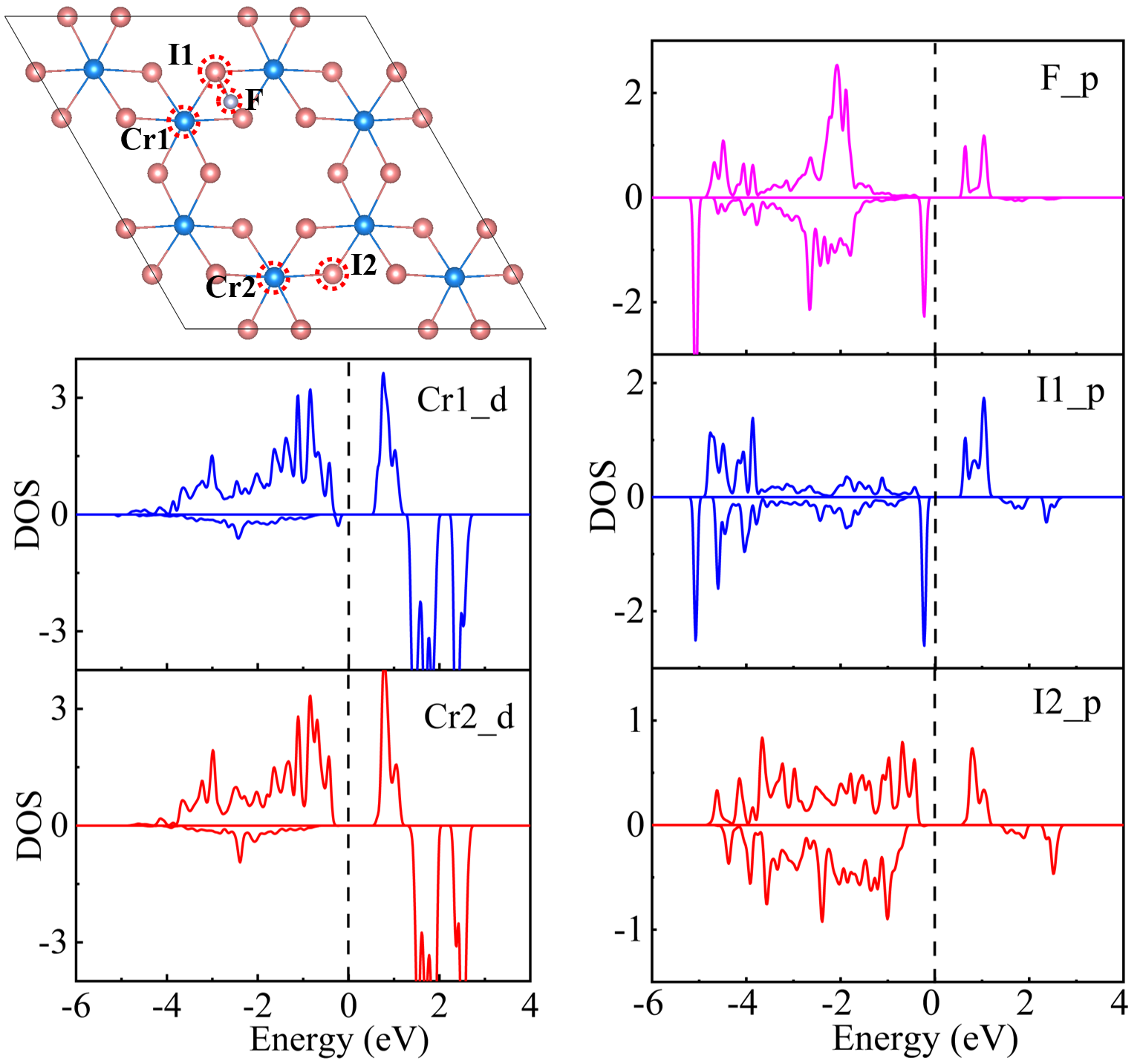}
\caption{The atomic projected density of states of Cr1, Cr2, I1, I2 and F atom as marked in the atomic structure diagram.}
\label{Fig3}
\end{center}
\end{figure}

To further understand the effects of X adatom on local electronic structures, we project the density of states to F adatom, I1, I2, Cr1 and Cr2 atoms in F@CrI$_3$, as illustrated in Fig.~\ref{Fig3}. The projected density of states (PDOS) of I2, Cr1 and Cr2 atoms do not display obvious difference from pristine CrI$_3$. It suggests the states of these atoms suffer negligible influence from X adatom, as they not bond with X adatom directly. The PDOS of F adatom in spin-up and spin-down directions are asymmetric, which is consistent with the above finding that F adatom is spin polarized with net magnetic moment. In the PDOS of F adatom and I1 atom, there are two intense resonant states located at -5.0 and -0.1 eV, confirming the strong hybridization between F and nearest neighbor I atom. However, the states of I1 and I2 demonstrate significant difference, for the intensity of occupied states of I1 from -3.8 to -0.2 eV is much weaker than I2, which reflecting different chemical environments. Therefore, by the adsorption of X atom, the states of ligand I atom that bonds with X adatom directly will be manipulated.

\begin{figure*}
\begin{center}
\includegraphics[width=0.94\textwidth]{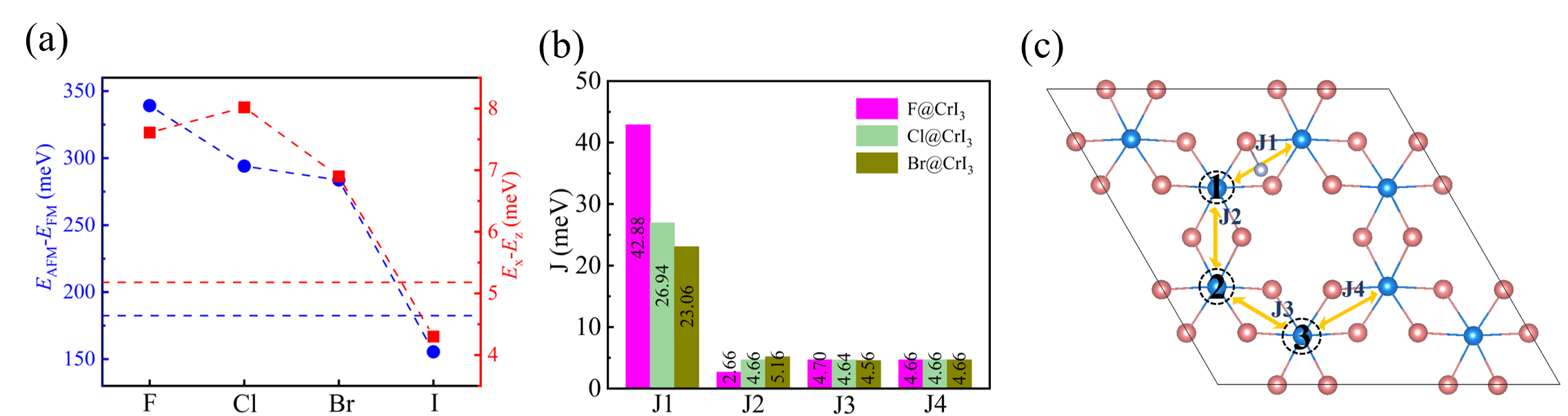}
\caption{The exchange energy \emph{E}$_{AFM}$-\emph{E}$_{FM}$ (blue) and magnetic anisotropy energy \emph{E}$_{x}$-\emph{E}$_z$ (red) of X@CrI$_3$. (b) The exchange interaction parameters in CrI$_3$@F as shown in (c).}
\label{Fig4}
\end{center}
\end{figure*}

In CrI$_3$ sheet, ligand I atom bonds with magnetic Cr atom and is spin polarized in antiparallel direction. Interestingly, the spin polarization of X adatom is parallel to the I atom, and antiparallel to Cr atom. Consequently, the total magnetic moment of CrI$_3$ supercell is reduced by 1 $\mu$$_B$. It should be noted that the calculation cannot achieve the state that the magnetic moment of X adatom is antiparallel to I (parallel to Cr), even if we set this state initially. The magnetic moment M$_I$ of I atom in pristine CrI$_3$ is -0.081 $\mu_B$. However, for the I atom that bonds with F adatom, this value increases to -0.377 $\mu_B$. This phenomenon may be due to the charge transfer from I atom to X adatom, which leaves behind unpaired electrons at I atom. On the other hand, the transferred electrons pair with the electrons of X atom and reduces the magnetic moment M$_X$. For example, the magnetic moment of F adatom is -0.29 $\mu$$_B$, which is small than the 1 $\mu$$_B$ of isolated F atom. To further extract the exchange splitting $\Delta$$^{\uparrow\downarrow}$ of I atom in CrI$_3$, we calculate the difference between on-site energies of spin-up and spin-down Wannier orbitals in local coordinates. In pristine CrI$_3$, $\Delta$$^{\uparrow\downarrow}$$_{px(y)}$=0.023 eV and $\Delta$$^{\uparrow\downarrow}$$_{pz}$=0.068 eV. However, for the I atom that bonded with F adatom in F@CrI$_3$, the $\Delta$$^{\uparrow\downarrow}$ for p$_x$, p$_y$ and p$_z$ are 0.560, 0.854 and 0.148 eV, which are increased by an order of magnitude. Thus the stability of the magnetism of ligand I atom in CrI$_3$ can be enhanced by X adatom.

As the MA is crucial in 2D magnetic material, we calculate the MA energy \emph{E}$_{MAE}$ by
\begin{equation}
E_{MAE}=E_x-E_z
\end{equation}

\noindent Where the \emph{E}$_x$ and \emph{E}$_z$ are the total energy of X@CrI$_3$ when magnetism direction points to x and z directions. The results are plotted in Fig.~\ref{Fig4}(a). We can see the \emph{E}$_{MAE}$ of X@CrI$_3$ (X=F, Cl and Br) are 7.61, 8.02 and 6.90 meV, which are larger the 5.18 meV of 2$\times$2 CrI$_3$. However, I adatom reduces the \emph{E}$_{MAE}$ to 4.30 meV.

To determine the ground magnetic order, the EX energy \emph{E}$_{EX}$ is calculated by
\begin{equation}
E_{EX}=E_{AFM}-E_{FM}
\end{equation}

\noindent Where \emph{E}$_{AFM}$ and \emph{E}$_{FM}$ are total energy of X@CrI$_3$ at Ne\'{e}l antiferromagnetic and ferromagnetic state. As we can see in Fig.~\ref{Fig4}(a), similar to \emph{E}$_{MAE}$, the \emph{E}$_{EX}$ of X@CrI$_3$ (X=F, Cl and Br) is much larger the pristine CrI$_3$. For example, the \emph{E}$_{EX}$ of F@CrI$_3$ is 339.1 meV, almost twice as much as the 182.4 meV of 2$\times$2 CrI$_3$. On the contrary, \emph{E}$_{EX}$ of I@CrI$_3$ is 155.4 meV, which is smaller than the pristine CrI$_3$. Upon the adsorption of X adatom, both atomic and electronic structures of CrI$_3$ change. In order to clarify which fact plays the primary role in the regulation of ferromagnetism in CrI$_3$, we calculate the \emph{E}$_{EX}$ and \emph{E}$_{MAE}$ of unrelaxed F@CrI$_3$, and they are 342.7 and 8.4 meV, respectively. On the contrary, if we remove the F adatom from the relaxed F@CrI$_3$, then the \emph{E}$_{EX}$ and \emph{E}$_{MAE}$ are calculated to be 183.5 and 5.0 meV, which are very close to pristine CrI$_3$. Hence the modulation of magnetism of CrI$_3$ that caused by adatom is mainly due to the change of electronic states.

In the above we find X adatoms modulate the states of the bonded I atom pronouncedly, while exert very little influence on other atoms. In monolayer CrI$_3$, the states of I ligands play crucial role in the formation of ferromagnetism.\cite{besbes2019microscopic} As a consequence, the ferromagnetism in CrI$_3$ may be changed locally. To confirm this, based on a ferromagnetic state, we flip the magnetic moment of one Cr atom and calculate the energy change $\Delta$\emph{E}. As illustrated in Fig.~\ref{Fig4}(d), the moments of Cr1, Cr2 and Cr3 are flipped successively. In Heisenberg model, the magnetic exchange energy can be expressed as $E=-\sum_{\langle ij \rangle} J\vec{S}_i\cdot\vec{S}_{j}$, where J is exchange coupling parameter and S=3/2 is magnetic quantum number of Cr atom.\cite{kittel1996introduction} We just consider the nearest neighbor coupling, then $\Delta E$ can be expressed approximatively as
$$
\left\{
\begin{array}{c}
\Delta E_1=2J_1S^2+4J_2S^2\\
\Delta E_2=2J_2S^2+4J_3S^2\\
\Delta E_3=2J_3S^2+4J_4S^2
\end{array}
\right.
$$

\noindent By this method, the J of pristine CrI$_3$ is calculated to be 4.66 meV. The $\Delta$\emph{E}$_3$ results from the flipping of Cr3 atom is 63.2 meV, and it is very similar to the 62.9 meV in pristine CrI$_3$. Consequently, J$_4$ is expect to approximate the nearest neighbor exchange interaction of pristine CrI$_3$. Then we can solve J$_1$, J$_2$ and J$_3$, and the results are plotted in Fig.~\ref{Fig4}(b). We find that only J$_1$ increases greatly, such as in F@CrI$_3$, J$_1$ increases to 42.88 meV, which is about ten times as large as the J$_4$. Nevertheless, the J$_2$ and J$_3$ are very close to J$_4$. Therefore, X adatom can enhance J$_1$ markedly, while has little influence on J$_2$, J$_3$ and J$_4$.

\begin{table*}
  \caption{The calculated hopping matrix element $\mid$t$_{pd}$$\mid$ and energy difference $\mid$U$_{pd}$$\mid$ between I p and Cr d orbitals in pristine CrI$_3$ and F@CrI$_3$. The Cr1, Cr2, I1 and I2 atoms are marked in Fig.~\ref{Fig3}. The unit is eV.}
  \label{tb2}
  \setlength{\tabcolsep}{6mm}{
  \begin{tabular}{cccccc}
    \hline
     \quad & \multicolumn{5}{c}{Hopping matrix element $\mid$t$_{pd}$$\mid$}\\
     \cline{2-6}
     & $d_{z^2}$-p$_z$ & $d_{yz}$-p$_z$ & $d_{xz}$-$p_z$ & $d_{x^2}$-$p_z$ & $d_{xy}$-$p_z$ \\
    \hline
    CrI$_3$ & 0.019260 & 0.027227 & 0.002542 & 0.034491 & 0.484550 \\
    F@CrI$_3$ (Cr1-I1) & 0.012661 & 0.048880 & 0.026073 & 0.065627 & 0.630948 \\
    F@CrI$_3$ (Cr2-I2) & 0.019112 & 0.025609 & 0.002415 & 0.036971 & 0.485436 \\
    F@CrI$_3$ (Cr1-F) & 0.059039 & 0.017436 & 0.018420 & 0.113458 & 0.043808 \\
      \hline
       \quad & \multicolumn{5}{c}{Energy difference $\mid$U$_{pd}$$\mid$}\\
     \cline{2-6}
     & $d_{z^2}$-p$_z$ & $d_{yz}$-p$_z$ & $d_{xz}$-$p_z$ & $d_{x^2}$-$p_z$ & $d_{xy}$-$p_z$ \\
     \hline
     CrI$_3$ & 0.9092 & 0.1290 & 0.1291 & 0.9103 & 0.1292 \\
    F@CrI$_3$ (Cr1-I1) & 0.0134 & 0.0448 & 0.7346 & 00448 & 0.0254 \\
    F@CrI$_3$ (Cr2-I2) & 0.8877 & 0.1144 & 0.1175 & 0.8881 & 0.1162 \\
    F@CrI$_3$ (Cr1-F) & 1.5013 & 1.4699 & 2.2493 & 1.4699 & 1.4893 \\
      \hline
  \end{tabular}}
\end{table*}

In CrI$_3$, the EX between Cr atoms is mediated by ligand I atoms. It is realized by the virtual hopping between Cr d orbitals and intermediate I p orbitals. As the bond angle of Cr-I-Cr is about $90^\circ$, the effective EX between Cr atoms is ferromagnetic. According to the superexchange theory, the magnitude of J of CrI$_3$ is direct proportion to $\frac{t_{pd}^4}{U_{pd}^3}$, where $t_{pd}$ and $U_{pd}$ are the hopping matrix element and energy difference between Cr d and I p orbitals, respectively.\cite{khomskii2014transition} Based on the Bloch wave functions obtained from density functional calculations, we construct the maximum localized Wannier orbitals in local coordinate. By these Wannier orbitals, the t$_{pd}$ and U$_{pd}$ in CrI$_3$ and F@CrI$_3$ are calculated, as summarized in Table~\ref{tb2}. We can find there is obvious increase in the t$_{pd}$ between Cr d orbitals and p orbitals of the I atom that bonded with F adatom. Conversely, the $\Delta_{pd}$ drops significantly. Finally, the effective exchange coupling J increases. On the contrary, the t$_{pd}$ and $U_{pd}$ in Cr-I pair that is far away from F adatom are very similar to pristine CrI$_3$, thus the J will not change obviously. This is consistent with the result derived from magnetic Hamiltonian in the above. We note that the hopping parameters between Cr d and F p orbitals are nonzero, and the magnitude can be up to 0.1 eV. As the bond angle Cr-F-Cr is $63^\circ$, according to GKA rule,\cite{gka} the X adatom may provide additional ferromagnetic coupling channel, and contributes to the ferromagnetism in CrI$_3$.

\subsection{\label{sec:level2}Bi-atom adsorption}
\begin{figure}
\begin{center}
\includegraphics[width=0.48\textwidth]{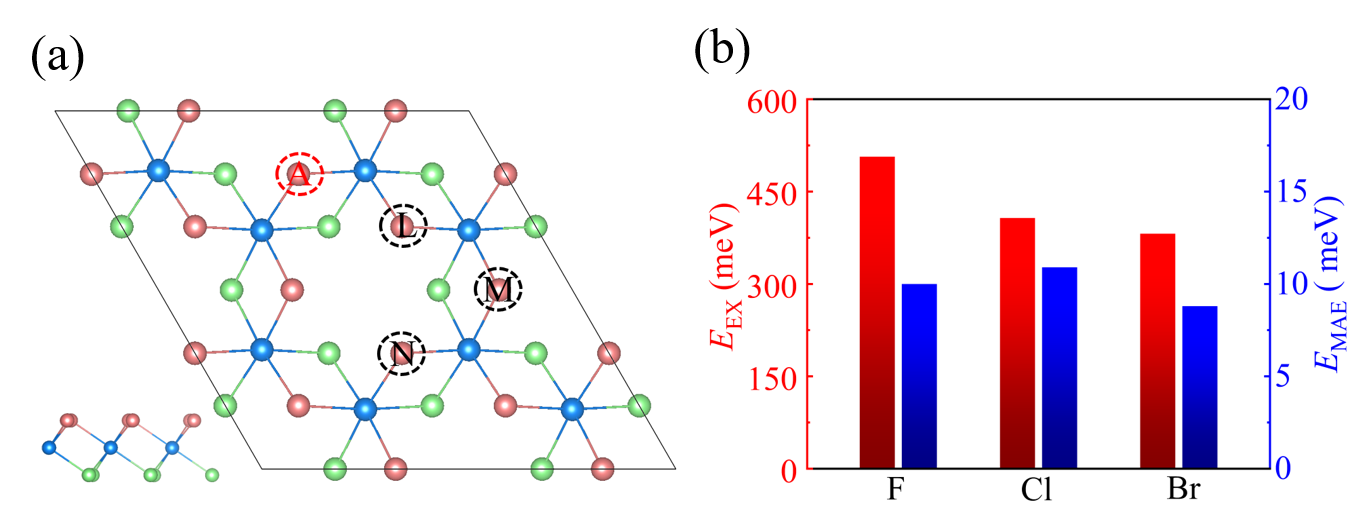}
\caption{(a) The possible adsorption configurations of X-dimmer. One atom at A site, the other one can be located at L, M or N site. (b) The E$_{EX}$ and E$_{MAE}$ of CrI$_3$ adsorbed X-dimmer at AN configuration. Red represents E$_{EX}$ and blue represents E$_{MAE}$.}
\label{Fig5}
\end{center}
\end{figure}

To explore effects of multiple X adatoms on the ferromagnetism of CrI$_3$, we adsorb two X atoms on 2$\times$2 supercell of CrI$_3$, and the associated coverage is 16.6\%. Different from single atom adsorption, the interaction among multiple adatoms may be developed. To understand the interaction, we consider three configurations by placing one atom at position A, and the second one at L, M or N site, as sketched in Fig.~\ref{Fig5}. The total energies of bi-X@CrI$_3$ at AN are lower than AL by 19.7, 48.5 and 44.5 meV for F, Cl and Br, respectively. This is due to the Coulomb repulsive interaction, for X adatoms are charged by charge transfer, and indicates X adatoms prefer to distribute uniformly on CrI$_3$ surface. Note that two I atoms on CrI$_3$ surface will desorb as I$_2$ molecule. The EX energy \emph{E}$_{EX}$ and MA energy \emph{E}$_{MAE}$ of bi-X@CrI$_3$ at most energetically favorable configuration are calculated, and depicted in Fig.~\ref{Fig5}. Compared with single atom adsorption, bi-adatom enhance the \emph{E}$_{EX}$ and \emph{E}$_{MAE}$ by greater extend. For instance, the \emph{E}$_{EX}$ and \emph{E}$_{MAE}$ of bi-F@CrI$_3$ are 499.2 meV and 11.03 meV, respectively, which are 2.74 and 2.13 times as much as the pristine CrI$_3$. This is because the X adatom increases CrI$_3$ ferromagnetism locally, and more adatoms result in more increased sites, as well as the greater total EX energy.

\subsection{\label{sec:level2}Fully F-functionalized CrI$_3$}
\begin{figure}
\begin{center}
\includegraphics[width=0.45\textwidth]{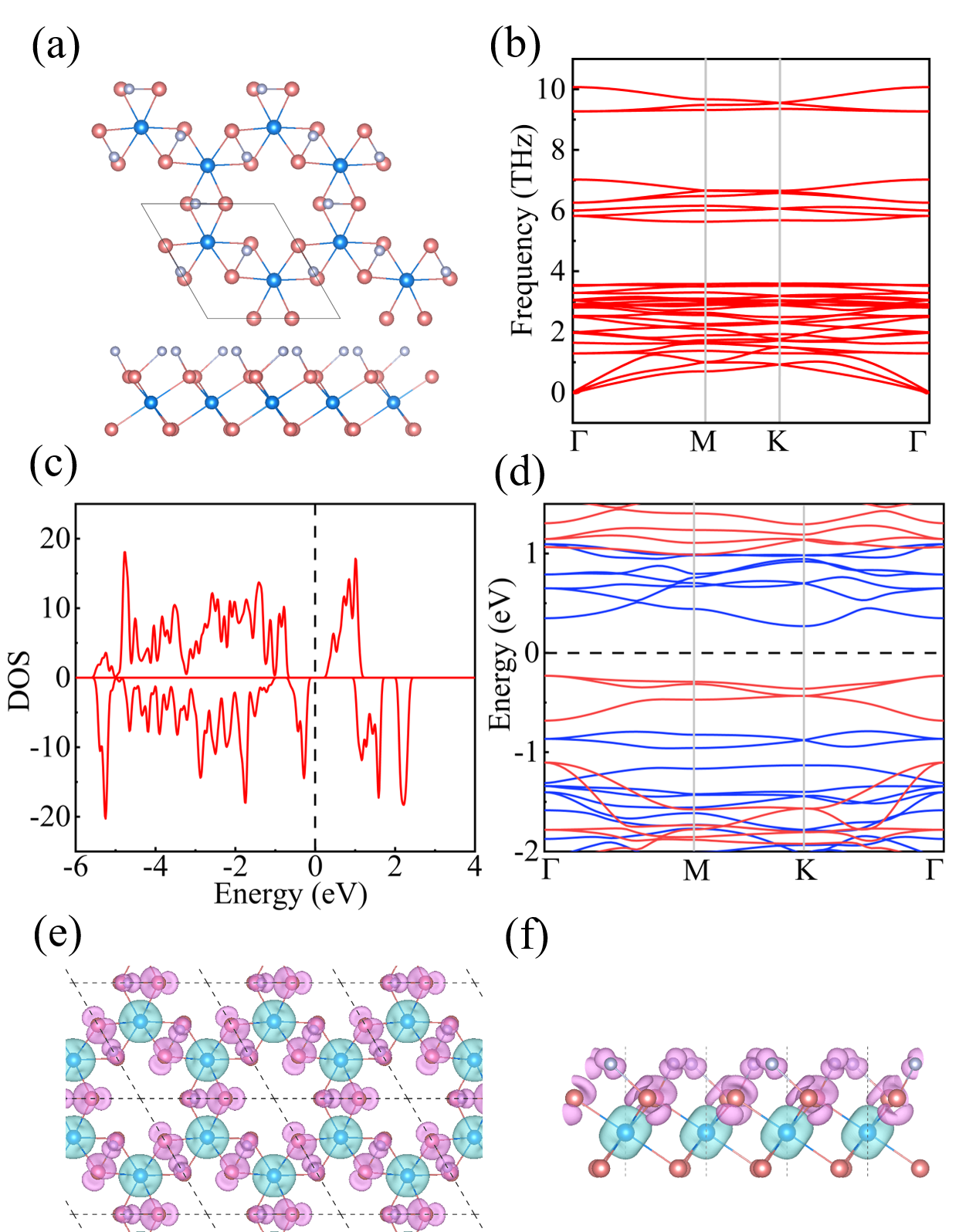}
\caption{(a) The atomic structure, (b) phonon spectrum, (c)(d) DOS and band structure, (e)(f) top and side view of spin charge density of FF@CrI$_3$. The isosurface is set to 0.012 e/{\AA}$^3$.}
\label{Fig6}
\end{center}
\end{figure}

Inspired by the EX and MA increase with coverage, finally we fully functionalize one side of CrI$_3$ by X atoms. However, only the system functionalized by F atoms (noted as FF@CrI$_3$) is found to be dynamic stable, as there is no imaginary frequency in whole Brillouin zone. Fig.~\ref{Fig6}(a) shows the top and side views of relaxed FF@CrI$_3$. The length of F-I bond is 2.17 {\AA}, which is shorter than the 2.20{\AA} in the adsorption of single F atom. On the other hand, the adsorption energy per F adatom is 2.585 eV, and it is larger than the 2.216 eV for single F adatom. Besides, the length of upper Cr-I bond is 2.683 {\AA}, while the bottom one is 2.767 {\AA}. The two values are smaller or larger than the 2.747 {\AA} in pristine CrI$_3$. Consequently, the lattice parameter of FF@CrI$_3$ is shrunk by 2\% relative to pristine CrI$_3$.

Fig.~\ref{Fig6}(c) and (d) shows the band structure and DOS of FF@CrI$_3$. The conduction band minimum and valance band maximum locate at K and $\Gamma$ point, respectively, leading to an indirect gap of 0.5 eV. By comparing with the band of CrI$_3$ adsorbed with single F adatom, we can conclude the valence band is formed by the localized states. Nevertheless, the valence band is delocalized by the interactions among localized states. We note that the valence band is spin-down, in contrast to the spin-up of conduction band. This is distinct from CrI$_3$, in which both conduction and valance bands are spin-up.

Now we would like to understand the magnetic properties of FF@CrI$_3$. Fig.~\ref{Fig6}(e) and (f) depict the magnetic moment distribution of FF@CrI$_3$. The Cr atom has 3 $\mu_B$ upward moment, which is the same as CrI$_3$. F adatom and the bonded I atom have 1 $\mu_B$ downward moment totally. As a result, the total moment of unit FF@CrI$_3$ reduces to 3 $\mu_B$. By taking account of the Heisenberg exchange coupling and single ion anisotropy,\cite{xiang2013magnetic} the magnetism of FF@CrI$_3$ can be described as

\begin{equation}
\hat{H}=-\sum_{\langle ij \rangle} J\vec{S_i}\cdot\vec{S_j}-\sum_i D{S_i^z}^2
\end{equation}

\noindent Where J is the EX parameter, D is the single-ion magnetic anisotropy parameter, $S_i^z$ represents projection of $\vec{S}$ along z direction, and the off-plane direction is chosen as z axis. By this definition, $J\textgreater0$ favors ferromagnetic interactions and $D\textgreater0$ favors off-plane easy axis. Because the occupation of Cr orbitals in FF@CrI$_3$ is very similar to CrI$_3$, the S is still equal to 3/2. We calculate the total energies at four different magnetic configurations, namely z-(anti)ferromagnetism and x-(anti)ferromagnetism. For CrI$_3$, the J and D are calculated to be 2.69 meV and 0.34 meV, respectively. By Metropolis Monte Carlo (MC) method, the T$_C$ is evaluated to be about 40 K, which is very similar to the experimental value and verifies the reliability of our calculation. For FF@CrI$_3$, the J and D are 55.05 meV and 0.55 meV,respectively, and the simulated T$_C$ is about 650 K. Therefore, FF@CrI$_3$ is a room temperature 2D ferromagnetic semiconductor.

\subsection{\label{sec:level2}Functionalization by other atoms}
In addition to halogen, we also study the adsorption of other nonmetal atoms on CrI$_3$, such as H, B, C, N, O, Si, P, S, As, and Se atoms. However, the Cr-I bond of CrI$_3$ is broken by the adsorption, and the left Cr or I atom will pair with the adatom. For example, H adatom bonds with I atom, and O adatom bonds with Cr atom. As a result, the structure of CrI$_3$ will be destroyed by these adatoms. Therefore, only halogen can enhance the ferromagnetism of CrI$_3$, as well as maintain its structure.

\section{\label{sec:level1}Conclusion}
In summary, based on first-principle calculations, a systematical investigation about the properties of CrI$_3$ adsorbed by X atom (X=F, Cl, Br and I) was carried out. We find both magnetic coupling and magnetic anisotropy of CrI$_3$ can be enhanced by X (X=F, Cl and Br) adatom, and further increase with the coverage of adatom. One side fully F-functionalized CrI$_3$ is a ferromagnetic semiconductor with T$_C$=650 K. Our study demonstrates that surface functionalization is a promising way to improve the ferromagnetism of 2D magnetic material, and finds a room temperature 2D ferromagnetic material.

\subsection*{Acknowledgments}
This work was supported by National Natural Science Foundation of China (grant No. 11874092), the Fok Ying-Tong Education Foundation, China (grant No. 161005), the Planned Science and Technology Project of Hunan Province (grant No. 2017RS3034), Hunan Provincial Natural Science Foundation of China (grant No. 2016JJ2001 and 2019JJ50636), Scientific Research Fund of Hunan Provincial Education Department (grant No. 18C0227) and Open Research Fund of Hunan Provincial Key Laboratory of Flexible Electronic Materials Genome Engineering (grant No.201905).

\bibliography{ref}
\end{document}